# Software-Defined Networking-based Crypto Ransomware Detection Using HTTP Traffic Characteristics

Krzysztof Cabaj[1], Marcin Gregorczyk[2] and Wojciech Mazurczyk[2]

[1]Warsaw University of Technology, Institute of Computer Science, Warsaw, Poland
[2]Warsaw University of Technology, Institute of Telecommunications, Warsaw, Poland
email: kcabaj@ii.pw.edu.pl, {M.Gregorczyk, wmazurczyk}@tele.pw.edu.pl

*Abstract* — Ransomware is currently the key threat for individual as well as corporate Internet users. Especially dangerous is crypto ransomware that encrypts important user data and it is only possible to recover it once a ransom has been paid. Therefore devising efficient and effective countermeasures is a rising necessity. In this paper we present a novel Software-Defined Networking (SDN) based detection approach that utilizes characteristics of ransomware communication. Based on the observation of network communication of two crypto ransomware families, namely CryptoWall and Locky we conclude that analysis of the HTTP messages' sequences and their respective content sizes is enough to detect such threats. We show feasibility of our approach by designing and evaluating the proof-of-concept SDN-based detection system. Experimental results confirm that the proposed approach is feasible and efficient.

## 1. Introduction

Year 2016 has been named by mass media as "the year of the ransomware" and this type of threat is currently considered by the security community and law enforcement agencies (see e.g. Europol's recent "2016 Internet Organized Crime Threat Assessment" report [7]) as a key threat to Internet users. Ransomware is a type of malicious software that is designed for the direct revenue generation and which after infection holds victim's machine or user's critical data "hostage" until a payment is made. Ransomware developers are constantly improving their "products" making it harder to design and develop effective and long-lasting countermeasures. Considering the fact that more and more devices is foreseen to be connected to the Internet due to e.g. Internet of Things (IoT) paradigm makes it a perfect environment for ransomware to spread in a foreseeable future [9]. The ransomware plague has been currently so widely sprawled that there are even crime-as-a-service tools available in the dark web (like TOX ransomware-construction kit [8]) which allow even inexperienced cybercriminals to create their own customized malware, to manage infections, and profits.

There are two main types of modern ransomware i.e. locker and crypto. The infection for both kinds of malicious software happens in the similar way i.e. a user machine is infected by means of various attack vectors, e.g., by drive-by-download, malvertisement, phising, spam, different forms of social engineering, etc. However, what comes after the infection is different for both types. *Locker ransomware* denies user access to an infected machine but typically the underlying system and files are left untouched. On the other hand, a *crypto ransomware* is a kind of a data locker that prevents the user from accessing her/his vital files or data (e.g., documents, pictures, videos, etc.) by using some form of encryption. Therefore attacked files are useless until a ransom is paid and the decryption key is obtained. Then after the user's machine is locked or data is encrypted the victim is presented with an extortion message. In many cases paying the ransom to the cybercriminal is the only way to get back access to the machine/data. The value of the requested ransom differs and is typically in range US$300-$700, and the favored payment currency is bitcoins [9]. It must be emphasized that not only individual users are currently targeted but also companies and institutions like hospitals, law enforcement agencies, etc. Clearly effective and efficient solutions to counter ransomware infections are desired.

Although the first cases of crypto ransomware have been known for more than 10 years (e.g. Trojan.Gpcoder) it must be emphasized that the recent plague of this type of malware is related to the improved design of the cybercriminals' "products". The main difference now is that crypto ransomware moved from custom or symmetric key to asymmetric key cryptography (Fig. 1). In this case, when the machine is infected, it contacts C&C (Command & Control) server through the multiple proxy servers (which are typically legitimate but hacked machines) to request a public encryption key. At C&C a pair of matching public-private keys is generated for each infection and the public key is returned to the compromised host (private key never leaves the C&C server). Then the public key is used to securely transfer session key in order to encrypt the chosen files which are deemed most important for the user. It is worth noting that if correctly implemented, asymmetric crypto ransomware is (practically) impossible to break. The most prominent ransomware, and one of the first to introduce asymmetric key cryptography, is CryptoWall 3.0, which was discovered at the beginning of 2015, later followed by others CryptoWall 4.0, Locky, etc.

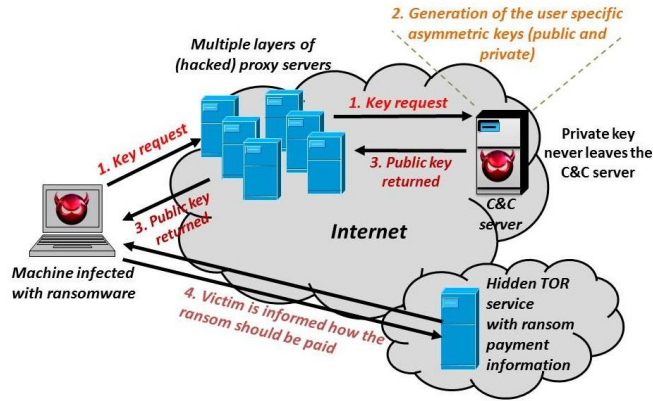

Fig. 1 Typical asymmetric key cryptography-based ransomware scheme.

Software-defined networking (SDN) is now one of the emerging networking paradigms [1]. Its main benefit is that it allows decoupling of the control and data planes i.e. the underlying network infrastructure is abstracted from the applications. Therefore the network can be managed in a logically centralized way. Apart from many potential applications [1] recently SDN has become also a promising opportunity to provide security for current networks in a more flexible and effective manner [2].

Taking above into account, in this paper we present a dedicated SDN-based system for ransomware detection and mitigation. It must be noted that when it is possible to successfully discover ransomware communication then, obviously, sometimes it may be too late to prevent encryption for that particular victim. However, it is still possible to utilize this incident to provide feedback for the detection system in order to "save" other users. This phenomenon is well-known in nature where a single organism often has to make self-sacrifice for the sake of the group [15]. As more and more analogies between cybersecurity and nature are continuously drawn [16] this may be another opportunity to reuse nature experience in this regard to improve communication networks' defenses.

The proposed in this paper detection system is focused on the crypto ransomware that utilizes asymmetric cryptography. While designing and developing the system we took into considerations results from the traffic analysis of two modern ransomware families i.e. CryptoWall and Locky. Based on the performed analyses we conclude that these ransomware families share some similarities which can be utilized to create an efficient and effective detection system. Thus, the main contributions of the paper are:
- Using the SDN-based architecture to mitigate crypto ransomware which allows to create a flexible and effective detection and prevention system,
- Presenting results of the network measurements-based behavioral analysis of two ransomware families namely CryptoWall and Locky,
- Designing the SDN-based detection and mitigation system that relies on our findings from the performed behavioral analyses mentioned above,
- Developing and evaluating the proof-of-concept implementation of the proposed detection system.

The rest of this paper is organized as follows. First, we present existing work related to malware detection using characteristics of the HTTP traffic, SDN-based approaches to cyber threats detection and prevention including ransomware, and in particular existing ransomware detection methods. In Section 3 we present results from the behavioral analyses of two ransomware families. In the next section, SDN-based ransomware detection system that relies on ransomware HTTP traffic characteristics is described and evaluated. Finally, Section 5 concludes our work.

## 2. Related work
In this section we review existing work first related to the ransomware detection, then SDN-based threat detection with special emphasis for ransomware countermeasures. Moreover, we analyse methods for discovering threats based on HTTP protocol traffic analysis.

Malware detection has been extensively studied in the recent years. In general such techniques can be broadly divided depending on where malware activities are observed either at the network-level ([29], [30]), system-level ([31], [32]), or both [33]. In this paper we propose a network-level SDN-based solution.

From the functioning perspective still the most common approach to malware detection is payload-based i.e. these approaches can be only effective if the malware communication is invariant [19]. Thus, if the malicious software is using plaintext communication protocol then such invariants may exist (e.g. protocol keywords can serve as a part of payload signatures). The same situation can be experienced also for several encrypted malware communication protocols, where certain invariant plaintext fragments result in certain invariant encrypted data (if encryption keys are not wisely used) which can be easily applied as signature as well [18]. However, often various malware families (including ransomware) do not exhibit characteristics indicated above making them able to circumvent payload signatures-based detection systems [20]. Recently a novel approach has been proposed which uses tamper resistant features of the transport layer protocol in order to distinguish malware heartbeat messages (that sustain the connection with C&C) from the legitimate traffic. However, authors noted that they observed substantial decrease in detection for malicious software which traffic is disguised in HTTP messages [21].

When it comes specifically to countering ransomware only several works exist. In [26] author propose Heldroid system which employs static taint analysis together with lightweight symbolic execution to find code paths that indicate device-locking activity or attempts to encrypt files on external media. Authors of [10] describe the results of a long-term study of ransomware between 2006 and 2014. Based on the gathered data they conclude that the number of ransomware families with sophisticated destructive capabilities remains quite small. They also propose detection system that is based on monitoring of abnormal file system activity. Scaife et al. [28] introduced an early-warning detection system for ransomware that monitors all file activities and alerts the user in case of something suspicious is identified using an union of three features i.e. file type changes, similarity measurement and entropy. Another recent work [27] introduced EldeRan which is a machine learning approach for dynamically analysing and classifying ransomware. The proposed solution monitors a set of actions performed by applications in their first phases of installation and tries to detect characteristics signs of ransomware. The obtained experimental results prove that this approach is effective and efficient showing also that dynamic analysis can support ransomware detection by utilizing the set of characteristic features at run-time that are common across families, and that helps the early detection of new variants.

When it comes to the SDN-based solutions tailored for security purposes the first work that proposed a general SDN-based anomaly detection system was put forward by Mehdi et al. in 2011 [2]. Authors showed how four traffic anomaly detection algorithms can be implemented in an SDN context using Openflow compliant switches and NOX controller. Obtained experimental results proved that these algorithms are significantly more accurate in identifying malicious activities in the home networks as compared to the ISP. Further, other researches utilized SDN also to detect network attacks [3] or to monitor dynamic cloud networks [5].

Several recently published papers deal also with SDN-based malware detection. Jin and Wang [4] analyzed malicious software behaviors on mobile devices. Based on the acquired knowledge they proposed several mobile malware detection algorithms, and implemented them using SDN. Their system was able to perform real-time traffic analysis and to detect malicious activities based only on connection establishment packets. In [25] authors designed and developed an SDN-based architecture specialized in malware analysis aimed to dynamically modify the network environment based on malicious software actions. They demonstrate that this approach is able to trigger more malware's events than traditional solutions. To the best of our knowledge no works so far proposed ransomware detection using SDN-based system.

HTTP traffic characteristics have been already used as a recognition property for detecting malware. For instance, a characteristic combination of HTTP URI request parameters in the malware communication may be utilized to discover malicious communication. For example, Perdisci et al. have shown that clustering HTTP traffic can be used to extract behavioral features which can be used to recognize HTTP-based malware [17]. In [22] Zegers investigated whether the order of HTTP request headers can be used to recognize malware communication. However, the author states that different websites has their own header order. Similar case is with the applications that communicate over HTTP (Windows updates and anti-viruses). Therefore the conclusion is that it is unfeasible to use HTTP headers order to reliably identify malicious software. Kheir [23] also analysed User Agent (UA) anomalies within malware HTTP traffic in order to extract signatures for its detection. His observation was that almost one malware out of eight uses a suspicious UA header in at least one HTTP request (this includes: typos, information leakage, outdated versions, and attack vectors such as XSS and SQL injection). Based on above the authors developed a new classification technique to discover malware user agent anomalies. Obtained experimental results showed that this approach considerably increases the malware detection rate. Similar approach was proposed in [24] where to detect malware authors take advantage of the statistical information about the usage of the UA of each user together with the usage of particular UA across the whole analysed network and typically visited domains.

However, it must be noted that so far no approach exists that takes into account the sequences of HTTP messages and their respective sizes as the main feature for the detection system as we propose in this paper.

The first dedicated SDN-based ransomware security solutions designed to improve the protection against the ransomware has been recently proposed in [14]. Authors introduced two approaches that take advantage of the fact that without successful communication to the C&C server the infected host is not able to retrieve the public key and, as a result, it cannot start the encryption process. Both methods rely on dynamic blacklisting of the proxy servers used to relay communication between an infected machine and the C&C server. However, the main drawback of the proposed approaches is that they are efficient only when the ransomware proxy servers were previously identified by means of behavioral analysis of known malware samples. Therefore it is not possible to discover new campaigns when they first appear.

Due to this fact, in the paper we take a different angle i.e. we utilize characteristics of the network communication between the infected host and a proxy server to detect ransomware data exchange. As we observe based on two "popular" ransomware families the communication protocol utilized in the analyzed malware samples is similar, therefore, utilization of this knowledge can lead to the development of an efficient and effective ransomware countermeasure. Such an approach can be also treated as a complementary solution to the detection methods proposed in [14] as it can form a source of feedback for the dynamic blacklisting of the proxy servers.

### 3. Crypto ransomware traffic characteristics based on CryptoWall and Locky families

Based on our experiences with crypto ransomware, we decided to choose two families namely CryptoWall and Locky to present their communication characteristics and to illustrate different levels of malware sophistication. These two ransomware examples utilize asymmetric cryptography, however, they differ when it comes to the network part of the communication although they both use HTTP protocol. That is why, in the following subsections we describe both ransomware families' communication characteristics in details. Table 1 summarizes our network measurement efforts for the two malware families indicated above.

| Ransomware type | No. of samples | No. of sample executions | Size of traffic traces |
|---|---|---|---|
| CryptoWall 3.0/4.0 | 359 (331 for CW3.0 and 28 for CW4.0) | 3 700 | ≈ 5 GB |
| Locky | 428 | 2 200 | ≈ 330 MB |

Table 1. Network measurement statistics for CryptoWall 4.0 and Locky ransomware.

### 3.1. CryptoWall communication characteristics

CryptoWall 4.0 has been active since October 2015 and it superseded the previous 3.0 version. From the network traffic perspective, to communicate with C&C server CryptoWall uses domain names instead of direct IP addresses. Analysis of the traffic from infected machines revealed that CryptoWall communication is based on HTTP POST messages. The communication is directed to the scripts uploaded onto the hacked web servers (proxy servers) and it is encrypted using the RC4 algorithm. However, the encryption key is very easy to retrieve as it is incorporated within the HTTP request.

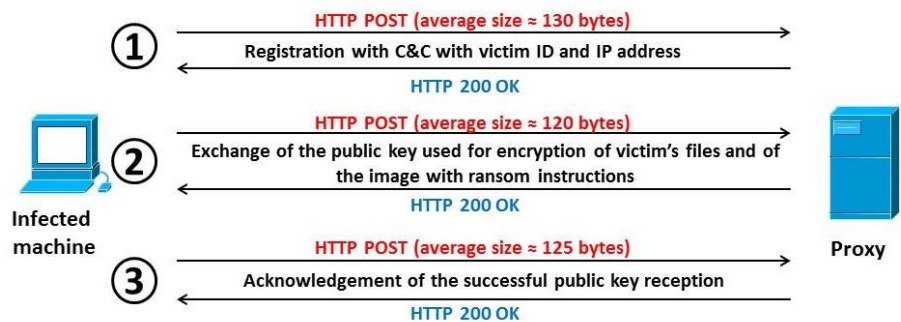

Fig. 2 CryptoWall 4.0 communication

The CryptoWall communication is depicted in Fig. 2. During the first data exchange, the ransomware reports its unique identifier and the victim's IP address to the C&C, which acknowledges the received information. In the second exchange, the response contains an image containing instructions for the victim and, TOR address of the ransom webpage, the victim's personal code, and an RSA 2048-bit public key that is used for encrypting the data. Then the encryption process starts. Finally, during the last data exchange an acknowledgement for the public key reception is provided. It is worth noting that

CryptoWall 4.0 is not reporting the finish of encryption process to the C&C as well as does not report the number of encrypted files, which was the case for its predecessor – CryptoWall 3.0.

During our research we investigated in total 359 CryptoWall samples, from which 28 were CryptoWall 4.0 and the rest CryptoWall 3.0. Each CryptoWall sample contained hardcoded list of proxy servers which is used during the transfer of public key from the attacker C&C server. We also discovered that typically these servers are victims, too. For executing proxy script cybercriminals behind CryptoWall are utilizing compromised legitimate servers. When a new campaign of CryptoWall appears there are many samples with the same proxy list. Our analyses for CryptoWall 4.0 revealed 19 proxy lists. The average proxy list contained on 47 servers' addresses (the shortest list had 27 and the longest 70 addresses). Using information concerning servers in proxy list we investigated how long these servers have been responsive. Fig. 3 and Fig. 4 illustrate the number of responsive servers in the detected proxy lists for CryptoWall 3.0 and 4.0 respectively. What should be emphasized the longest responding proxy server was active even as long as 11 weeks.

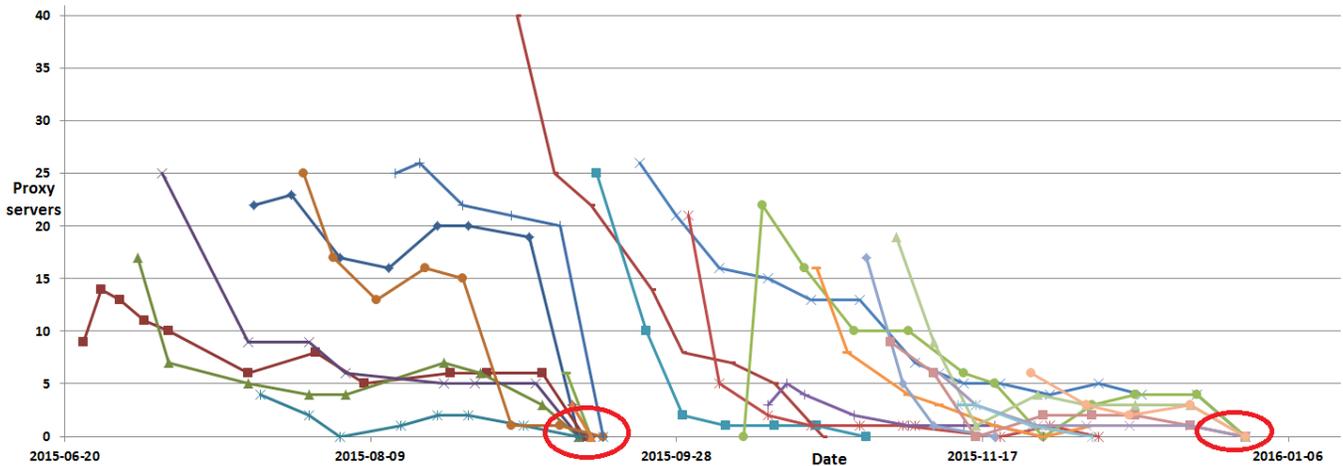

Fig. 3 CryptoWall 3.0 proxy servers' activity

Two time instants observed in Fig. 3 are particularly interesting and are highlighted with red ellipses. The first one is related to the massive shutdowns of proxy servers and immediate appearance of few new samples with new proxy lists. The second represents the complete shutdown of CryptoWall 3.0 infrastructure.

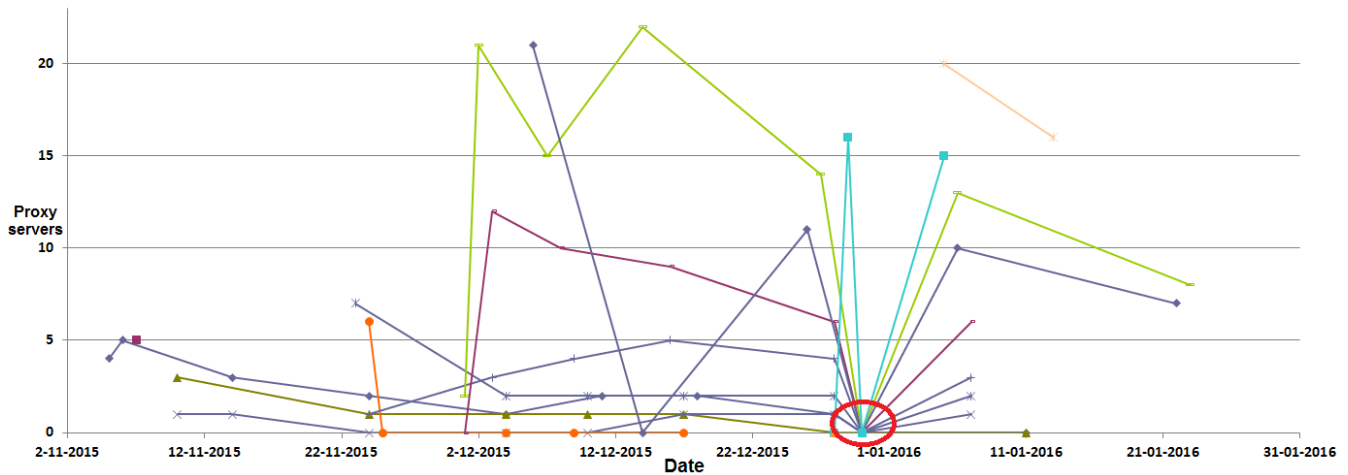

Fig. 4 CryptoWall 4.0 proxy servers' activity

It is worth noting that at the time of CryptoWall 3.0 shutdown there was a noticeable lack of CryptoWall 4.0 servers activity (Fig. 4). It might be assumed that during this time cybercriminals were performing (most probably) some kind of infrastructure update or maintenance. Another, more general conclusion is that typically the number of active servers from the proxy server list is decreasing in time.

### 3.2. Locky communication characteristics

First Locky ransomware samples have been observed since mid-February 2016. Its communication patterns are similar when compared with CryptoWall family. The first resemblance is related to utilization of HTTP POST messages. However, in this case the encryption scheme used to secure data exchanged has been chosen better. Thus while this article has been written, it was not possible to decrypt Locky communication. Therefore, information presented in this subsection is mostly deduced based on our previous experiences and knowledge about ransomware. The communication in case of Locky consists of four HTTP POST exchanges which (most probably) serve the same aim as in case of CryptoWall 3.0/4.0 (Fig. 2). For example, it is possible that the image transfer as well as the public encryption key transmission is performed during the third HTTP data exchange as the size of the resulting instructions presented to the user when extorting ransom is of similar size as the HTTP 200 OK response.

The other difference between Locky and CryptoWall is that disruption of the communication when transmitting the public key in case of CryptoWall was able to block the encryption process. However Locky in such cases did not stop encryption and used hardcoded key (probably one dedicated key per sample). . Additionally, in Locky the communication capabilities to contact C&C server have been significantly extended. Each Locky sample has its own list of C&C IP addresses hardcoded. In case these addresses are already blocked it uses DGA (Domain Generation Algorithm) to generate new C&C domains and tries to contact them. It is also worth noting that the first Locky samples generated exactly the same HTTP POST message sizes. However, Locky has evolved in time and currently the messages sizes are varying, as presented in Fig. 5.

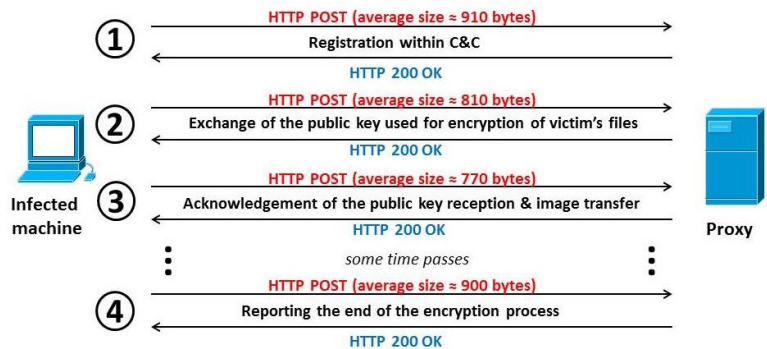

Fig. 5 Locky (most probable) communication.

We have started analyzing Locky since the end of March 2016. During these more than 8 months we observed various changes in Locky activities. The most interesting findings are provided further in this section. Until now, we observed three distinct versions of the Locky communication protocol. The first version used the same URL i.e. main.php and in all executions the sizes of the messages sent to the C&C server were the same: 101, 55 and 94 bytes of (probably) encrypted binary data. At the beginning of April the protocol has been changed, and then each execution of ransomware sample resulted in the randomly generated messages sizes. This behavior is similar to the change observed in CryptoWall family during transition from 3.0 to 4.0 version. The last observed modification to the communication protocol happened around the end of July, when the encoding of data using only ASCII printable characters has been first observed. This Locky version has greater messages' sizes and was used during our experiments.

From March 2016 we also observed changes in how the malware is distributed. Initially, during the first stage typically Microsoft Word or Excel document with embedded hostile macro was send in SPAM and later main stage of ransomware is downloaded from the distribution server and executed. Later, around the end of May, we started observing encrypted executables. The first stage macro in document downloaded an encrypted file, decrypted it on the infected machine and executed. Obviously, the file hosted on the distribution server without the decryption key cannot be executed. The number of distribution servers between March and November 2016 is illustrated in Fig. 6. It can be observed that till June 2016 the number of the distribution servers discovered daily was typically not higher than 50, however, in the last months it doubled with few spikes reaching more than 150. This means that Locky is still active and cybercriminals behind it still try to reach as many victims as possible. At the time of finishing of this article, after a week of silence, first sample of new campaign was discovered and analyzed.

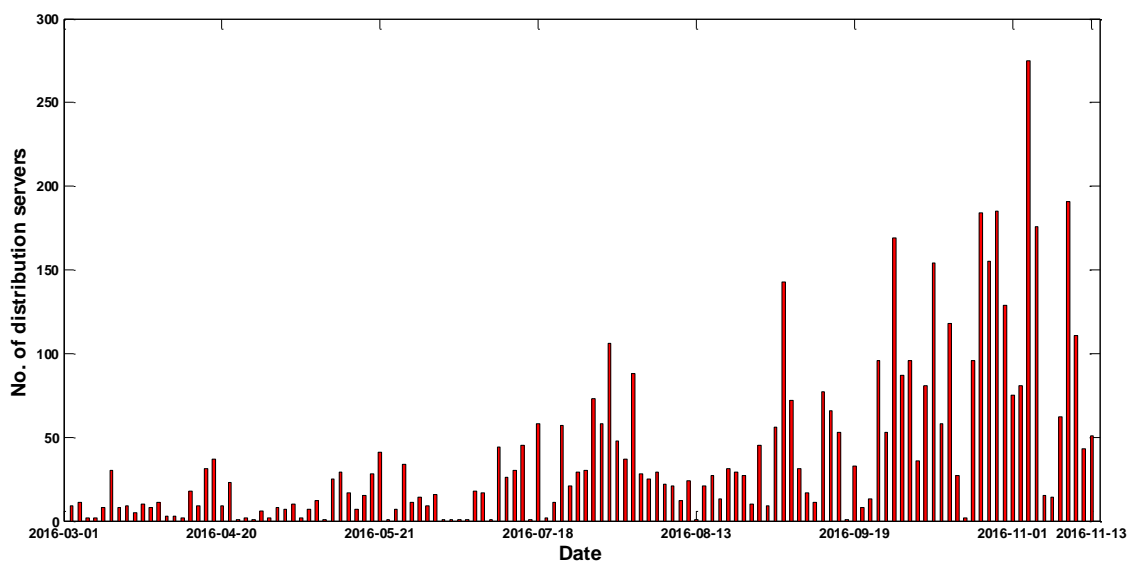

Fig. 6 Locky distribution servers' statistics (based on the data from http://ransomwaretracker.abuse.ch).

The most eye-catching aspect of the Locky are URLs used for C&C communication. During our research we investigated 11 distinct URLs. We decided to use their names as indicators of the new Locky campaigns. Our analyses revealed that each campaign lasted no longer than one month. From version 3.0 of the communication protocol we have observed various features of the gathered samples in more details. We discovered, among others, hardcoded C&C servers, used DGAs (Domain Generation Algorithms) and entry function's DLL names. All analyzed details are summarized in Table 2. Provided information allows illustrating the extent of resources used by cybercriminals' infrastructure.

| C&C Server URL | Date of the first sample analysis | No. of samples | No. of hardcoded C&C | No. of different DGA algorithms | DLL entry point name |
|---|---|---|---|---|---|
| main.php | 2016.03.21 | 41 | 18 | 7 | - |
| submit.php | 2016.03.28 | 24 | 15 | 3 | - |
| userinfo.php | 2016.05.03 | 226 | 42 | 9 | - |
| access.cgi | 2016.05.30 | 2 | 2 | 1 | - |
| /upload/_dispatch.php | 2016.05.31 | 18 | 14 | 9 | - |
| /php/upload.php | 2016.08.01 | 11 | 16 | 3 | - |
| /data/info.php | 2016.08.29 | 19 | 14 | 8 | 1 |
| apache_handler.php | 2016.09.27 | 20 | 22 | 8 | 1 |
| linuxsucks.php | 2016.10.24 | 9 | 11 | 5 | 2 |
| message.php | 2016.11.03 | 57 | 24 | 20 | 11 |
| information.cgi | 2016.11.21 | 1 | 3 | 1 | 1 |

Table 2. Statistics of the analyzed Locky samples.

The second column of Table 2 contains the date when we analyzed the first sample from a given campaign. We do our best to analyze samples as they appear, however, sometimes when attackers introduce more modification to their code it takes more time to analyze how the samples can be gathered. Therefore, the provided dates can be used only as an approximated time of the start of a new campaign. The fourth and fifth columns are associated with various aspects of communication with C&C server. As it was mentioned at the beginning of this section Locky uses more reliable C&C communication when compared to the CryptoWall. Firstly, Locky tries to contact hardcoded IP addresses of C&C servers. Our analysis shows that each sample has hardcoded from two to five such addresses within its binary. During our research we observed that if the hardcoded C&C servers are shut down then the next samples are equipped with previously unseen IP addresses of malicious servers. Due to this fact we observed averagely almost 18 C&C per analyzed campaign (maximum 42). In case when all hardcoded C&C servers are not active Locky switches to the second method: it utilizes DGA. What is interesting, samples from the same campaign uses various DGA algorithms, which in effect leads to various domains generated in during the same time frame. The number of used DGA algorithms is presented in the fifth column in Table 2. The sixth column contains the number of used Entry Functions. The first six campaigns used normal executables during the main stage of the ransomware infection. However, at the end of August we noticed a change in the malware distribution technique. We discovered that downloaded file was a DLL library. To execute malware in such form additional information i.e. the name of the entry function is needed. The first two campaigns utilizing such solution used the same, simple entry function with the name "qwerty". However, the next campaigns switched to more complicated and often modified entry function names. Obviously all such changes are introduced in order to make ransomware analysis more difficult and to increase the time needed for security professionals to understand its behavior.

## 4. SDN-based ransomware detection based on the HTTP traffic characteristics

As it was mentioned in the previous section all analyzed ransomware families utilize a custom protocol which is used for downloading encryption key, image, etc. from the attacker's Command & Control server. Due the fact that communication process is similar for both families this characteristic feature can be used for ransomware infections detection. That is why we use this knowledge to introduce a novel network traffic classification method which is based solely on the size of the data inserted by the victim into outgoing sequence of HTTP POST messages.

### 4.1. The proposed detection method

The proposed scheme can be divided into three main phases (Fig 7). The first is a learning phase in which using real ransomware samples and the network traffic generated by them we extract characteristic features from the outgoing HTTP messages (particularly their size). During the second, fine-tuning phase we focus on adjusting parameters of the detection method. Finally in the last, detection phase we utilize data gathered during the two previous phases and detect infections using the proposed SDN based solution. Later in subsection 4.4 we present results of the evaluation of the introduced detection method using both malicious and benign network traffic. It must be also emphasized that for each ransomware family the described three-phase procedure is conducted separately. All the phases are described in details in the following.

**Learning phase:** In this phase, data used later for the detection purposes are prepared and preprocessed. During this process ransomware samples for each ransomware family ($f$) that were accumulated (details on how this dataset was formed can be found in section 4.3) are executed in the controlled environment and all traffic generated by the infected machine is captured for further preprocessing. For this purpose we utilize Maltester environment as described in the [11]. From the obtained traffic first three HTTP POST messages sent by malware are located and the respective content lengths are extracted. In result, $k$th infection ($k \in (1, n)$, where $n$ is a total number of all ransomware samples for a given family), is characterized by a vector $S_{fk}$ consisting of three numbers [$s_{1k}$, $s_{2k}$, $s_{3k}$], each representing content size sent in the consecutive HTTP POST messages (POST triples vector). The set of all $S_{fk}$ is treated as a base to establish the main distinguishing feature for detection purposes.

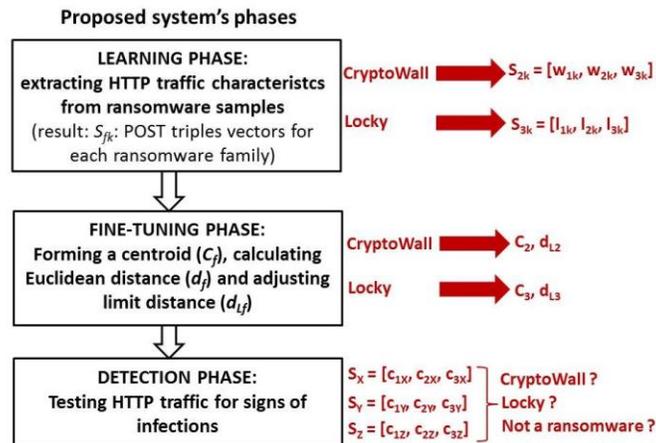

Fig. 7 The proposed detection method phases.

**Fine-tuning phase:** When the learning set of vectors $S_{fk}$ is (separately) prepared for each ransomware family ($f$) then the centroid vector of the corresponding feature vectors ($C_f$) as well as the minimal and maximal Euclidean distance for all vectors from $S_{fk}$ to $C_f$ are calculated. Then using $C_f$ and datasets that consist of the HTTP traffic from new infections (not utilized during the learning phase) and the benign HTTP traffic (without infections) the limit distance ($d_{Lf}$) from the centroid value is fine-tuned (the exact fine-tuning procedure is explained later in the experimental section). During fine-tuning phase information concerning minimal and maximal distance to centroid is utilized. As a result, at the end of this process all necessary parameters used later during the detection phase are calculated. In particular these are: three values to form the centroid vector $C_f$ and the limit distance $d_{Lf}$ that is used to assess whether currently evaluated feature vector extracted from the monitored HTTP traffic is malicious or not (Fig. 8, left). It must be also emphasized that for each ransomware family there will be a separate centroid vector $C_f$ as well as the limit distance $d_{Lf}$ established. All details concerning obtained experimental results are presented later in section 4.4.

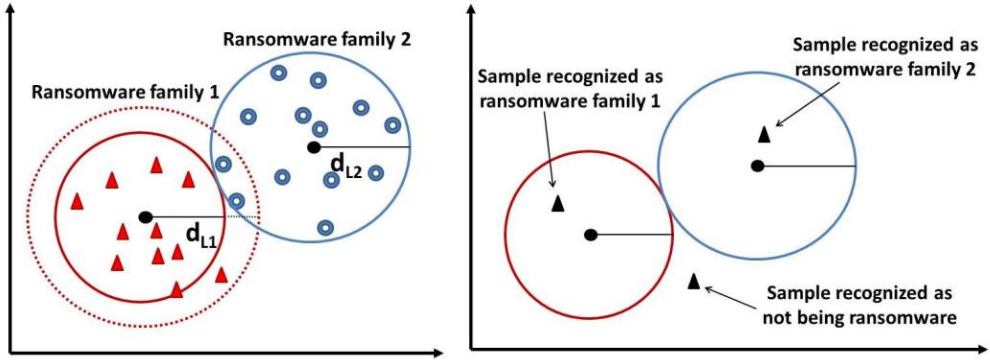

Fig. 8 Proposed detection method fine-tuning phase (left), detection phase (right)

**Detection phase:** HTTP connections that are subjected to monitoring and potential ransomware detection must be initially preprocessed. During our experiments we developed two variants of the preprocessing software. The first one, mainly for experimental use, extracts data from the traffic traces (files in the .pcap format). The second one is directly integrated within the SDN controller and can be used for real-time detection of the existing ransomware infections. Both preprocessing solutions analyze incoming TCP segments which contain HTTP traffic and reassembly outgoing messages. When the whole request is reassembled, then the size of the data sent to the server and host IP or domain address are extracted from the HTTP header. The extracted host IP or domain name is then used for defining destination HTTP server, which possibly could be a ransomware C&C or a proxy server. If confirmed as being malicious, information about such servers can be used, for example, to feed the ransomware countermeasures based on IP/domain blacklisting like the one proposed in [14] or directly block such traffic using the SDN-based system proposed in this paper.

In the next step, for each HTTP server identified as indicated above (malicious or not) a list is created that consists of the generated HTTP messages sizes. Any new HTTP message is thus added at the end of the list associated with a given server's IP address or a domain name. If after insertion of a new value to the list, its size is equal or greater than three (as for the proposed detection method we analyze 3 consecutive HTTP POST messages' sizes), then the detection of possible ransomware is performed. To this aim the test vector is constructed using last three values from the list. Then the distance between this vector and the centroid for a given ransomware family is measured. If the calculated value is smaller than the limit distance ($d_{Lf}$) established during the learning phase the system decides that this HTTP messages sequence is a sign of ransomware infection (Fig. 8, right). For example, consider HTTP POST messages' sequence presented in the Fig. 5 for Locky. In this case the extracted test vector is [910, 810, 770]. If during the learning phase the obtained centroid vector for this ransomware family has a value of [900, 800, 775] and the limit distance is fine-tuned to the value 20 then this traffic will be detected as ransomware, because the calculated Euclidean distance for this sample is 15.

### 4.2. Experimental test-bed

In the experimental evaluation we decided to utilize an Openflow protocol that enables typical networking devices to be supervised by an external controller and that is currently the standard for implementing SDN paradigm. The experimental environment used during experiments is depicted in Fig. 9. All operating systems were installed on virtual machines, which were managed by the VMware ESXi hypervisor installed at a server with Intel Xeon E5-2450 1.9GHz (24 logical CPUs, 24 GB RAM). In order to exclude any interfering factors, during experiments, no other virtual machines were active on the hypervisor.

In order to securely execute malware samples during experiments with SDN detection application Cuckoo Sandbox [12] has been used. In more details we utilized:
- Cuckoo guest with Microsoft Windows 7 SP1 x86_64 installed. We decided to use this OS, as currently it is dominant with almost 50% of desktops market share [13]. Therefore, it is likely that it will be targeted by cybercriminals. It must be also noted that the update and firewall services were intentionally disabled on this host so malware samples could be executed with no interruption.
- Pox/Cuckoo and Open vSwitch (OVS) host using Debian 8.5 x86_64 with kernel 3.16.36.

Due to separation purposes additionally two subnetworks were used: first for the Pox/Cuckoo traffic (using vNICs connected to Open vSwitch), and second for the management traffic (vNICs connected to VMware vSwitch).

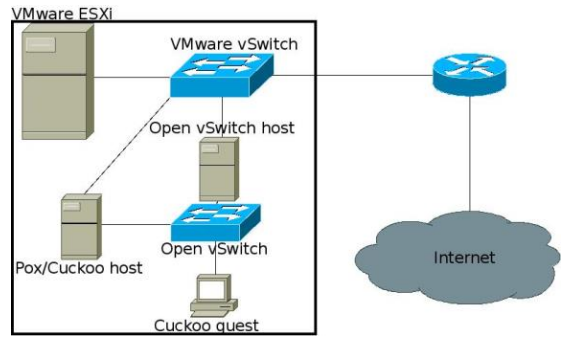

Fig. 9 Experimental testbed.

The functioning of the proposed test-bed was as follows. After Cuckoo agent (listener) was started on the guest machine, a snapshot with memory was taken. OVS was acting as a default gateway for Cuckoo guest, so the traffic directed towards the Internet was always passing through it. OVS was controlled using OpenFlow protocol by Python-based SDN controller – Pox. SDN application (based on the Layer 2 learning switch) forced OVS to forward every HTTP packet to the controller for inspection. Next, at the controller the decision on the packet was made. During experiments Cuckoo host first contacted ESXi API in order to restore Windows host from snapshot. Then ransomware samples were executed. In consequence the traffic directed towards the C&C server was generated based on which SDN detection and prevention system made decision whether to block it or not.

As stated before, Pox application is based on Layer 2 learning switch. First requisite was to configure OVS to send whole packet to SDN controller instead of default 120 bytes. It was needed as decision about rejecting or allowing traffic was made based on HTTP POST message size. Next, for every event (incoming packet) handler function was invoked to check if:
- packet is valid (can be parsed),
- TCP is used in the transport layer,
- TCP destination port is 80 (HTTP traffic),
- packet payload length is greater than 0 (to exclude TCP segments with no data, for example segments only with ACK flag).

If any of the checks failed, no blocking rule was set as such packet was not considered as malware traffic. If all checks were positive, payload was passed to the custom "oracle" function where the introduced detection algorithm was implemented (see section 4.1). If a ransomware infection is detected then the appropriate blocking flows (bidirectional using only hostile C&C server IP address) are inserted into OVS switch (otherwise the infection on any other machine cannot be not prevented).

### 4.3. Utilized datasets
This section describes how the relevant malicious and benign network traffic was gathered. We collected malware samples from two main locations: *malwr.com* website from which it is possible to directly download ransomware samples and *ransomtracker.abuse.ch* where one can obtain addresses of malicious distribution servers (the ones from which the ransomware binary is downloaded to the attacked host). Gathered samples are then executed in the controlled environments: the one described in section 4.2 and Maltester [11] where the generated test traffic was captured. This formed the first dataset with malicious HTTP traffic.

In order to confirm that proposed detection method is valid and does not generate too many false-positives, we utilized normal, benign HTTP traffic. As our proposed detection approach relies on HTTP POST messages characteristics then we needed representative network dumps. However, it must be noted that it is difficult to obtain unencrypted HTTP traffic captures as they might contain sensitive and private data (logins, passwords or cookies) thus such network dumps are usually not publically shared. Typically these repositories contain only metadata without upper layer payloads. Therefore the second dataset was generated using 200 000 of the most popular websites from Alexa ranking (extracted from

http://s3.amazonaws.com/alexa-static/top-1m.csv.zip on 19.10.2016). We wrote a script that launched each of the website in the Google Chrome web browser. As every website needs a certain amount of time to be fully loaded thus in order to speed up the process 25 pages were launched in the parallel for 25 seconds. Therefore the whole process of benign HTTP traffic generation took almost 56 hours but as a result we obtained real-life HTTP traffic as it is typically generated by Internet users.

The last part of the HTTP traffic was obtained from 2012 MACCDC (maccdc.org) cyber defense competition data dumps. It is a capture the flag type event which means that the HTTP data are not fully realistic, but they provided also examples of various types of network attacks and contained many POST messages. Table 3 summarizes the statistics of the gathered HTTP POST and domains values.

| Dataset name | No. of HTTP POST | No. of POST triples | No. of domains |
|---|---|---|---|
| Ransomware traffic for CW 4.0 | 750 | 250 | 250 |
| Ransomware traffic for Locky | 750 | 250 | 250 |
| Alexa traffic | 22 579 | 11 950 | 8 187 |
| MACCDC traffic | 17 249 | 15 862 | 761 |

Table 3. Characteristics of the datasets used for evaluation of the proposed detection method.

### 4.4. Experimental results
This section describes in details the process of the proposed detection method fine-tuning and presents obtained experimental results. The experiments were performed independently for CryptoWall 4.0 and Locky ransomware families. The following subsections describe results for Locky and CryptoWall families respectively.

#### 4.4.1 Locky ransomware
Initial data for the Locky family consists of more than 250 traces of successful infections with recorded data exchange between the infected victim and the C&C server. The complete dataset was divided into two distinct parts: the first which contains 150 samples that were used during learning phase for the centroid parameters and the limit distance calculations. For each point in this dataset, the distance to the centroid was calculated. The results of the average, minimal and maximum distance values were then used for determining the range of acceptable values for the limit distance parameters. For performance reason all calculations of the distance use distance square, which avoid CPU intensive calculation of the square root. The results for the learning phase include the minimal distance (actually distance square) of 6 914 and maximal of 274 370.

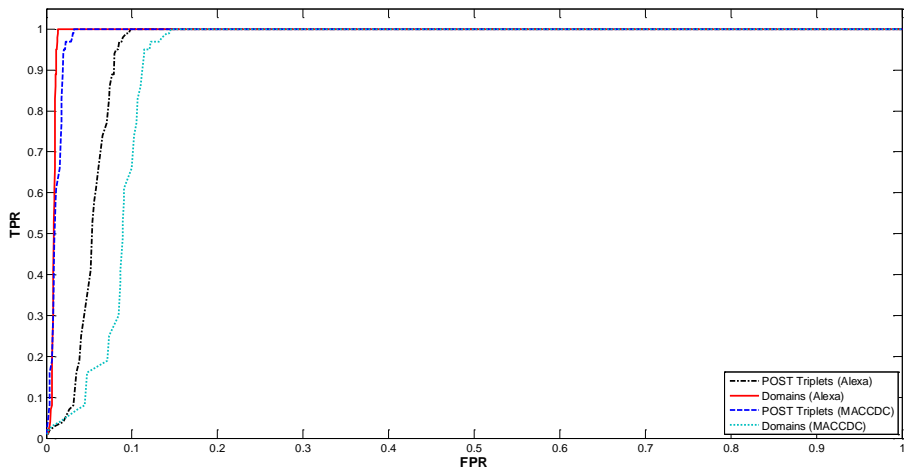

Fig. 10 The ROC curves for Locky detection for Alexa and MACCDC datasets.

The second part of the original dataset consists of 100 samples and it was used for the calculation of TPR (True Positive Rate) which was needed in order to prepare the ROC curve for the proposed detection method. This is a reasonable decision as this data was generated from the real infections (performed using the controlled and protected environment) and we are confident that they represent a real characteristic of Locky communication traffic. During this phase the number of traces flagged as containing Locky infections is analyzed for various i.e. average, minimal and maximum distance values. Due to fact that all data contains infections in the ideal situation we should flagged all of them as malicious. TPR is calculated as a ratio of flagged traces and the total number of samples in the dataset. The process of FPR (False Positive Rate) calculation is similar. We obtained clean traces using two distinct datasets, described in section 4.3, and we later reference to them as Alexa and MACCDC. In these datasets, contrary to the previous dataset which contained ransomware traffic, we should not detect any Locky infections. Therefore all connections recognized as Locky are false positives. For the ROC curve preparation purposes we checked these datasets using the same distance range as used for TPR calculations. During our research we investigated two types of ROC curves. The first one is associated with all tested consequent triples of POST messages. The second one is associated with domains, for which at least one triple is flagged as Locky transmission. Fig. 10 presents ROC curves independently for Alexa and MACCDC datasets and both features i.e. POST triples and domains.

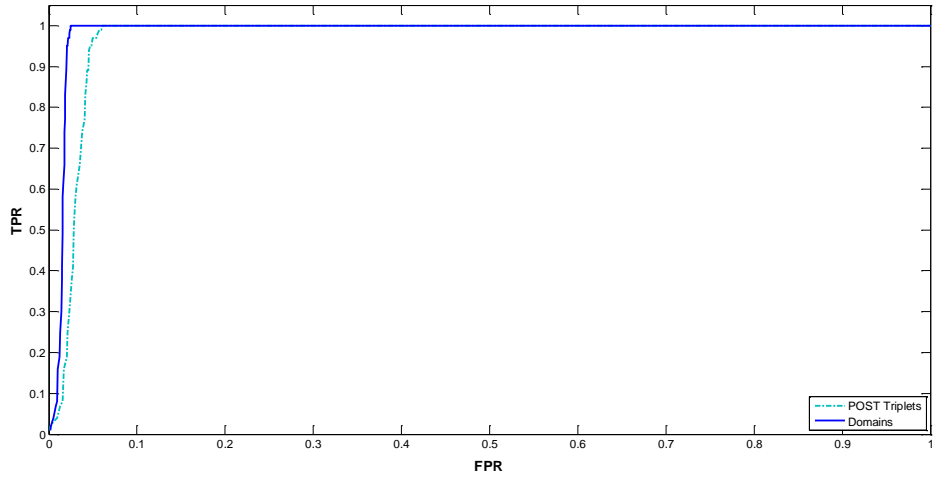

Fig. 11 The ROC curves for Locky detection for merged Alexa and MACCDC datasets and after final fine-tuning of the detector.

ROC plots for both datasets are similar, so we decided to merge them for the final fine-tuning of Locky detector. The final ROC curve is presented in Fig. 11. The results show that the most suitable limit distance ($d_L$) for our method is 260 000, because it offers the best trade-off between true positives (97%) and false positives (4.95%) and (2.2%) respectively for analyzed POST triples and domains.

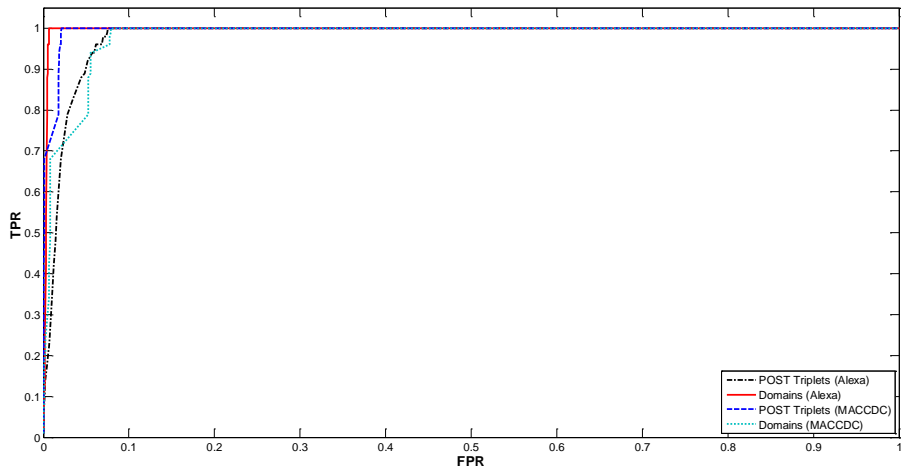

Fig. 12 The ROC curves for CryptoWall 4.0 detection for Alexa (left) and MACCDC (right) datasets.

### 4.4.2 CryptoWall 4.0 ransomware
During our research we performed similar analyses as presented in the subsection 4.4.1 also for CryptoWall 4.0. From our previous studies ([11], [14]) we have obtained almost 270 traces containing communication traffic between the infected host and CryptoWall C&C servers. Because this value is very similar to the number of traces used for Locky experiments we decided to split this dataset in a similar way i.e. to use 150 traces for the learning phase and 100 for the testing phase. It turned out that HTTP messages sizes sent by CryptoWall 4.0 are much smaller than in case of Locky family. After centroid calculation we discovered that the minimal distance (for CPU performance reasons it is actually distance square) is 25 and the maximal distance is 893. These values were then used during TPR and FPR calculations. For TPR, to obtain 100% detection accuracy we used limit distance of 1050. Using the second part of the traces containing CryptoWall C&C communications, Alexa and MACCDC datasets we gathered data for ROC curves which are illustrated in Fig. 12.

The ROC curves for both datasets are similar, so we decided to merge them for the final fine-tuning of CryptoWall detector. The final ROC plot is illustrated in Fig. 13. The results show that the best limit distance ($d_L$) is 900, because it offers the best trade-off between true positives – 98% and false positives – 4.2% and 1.2% respectively for analyzed the POST triples and domains.

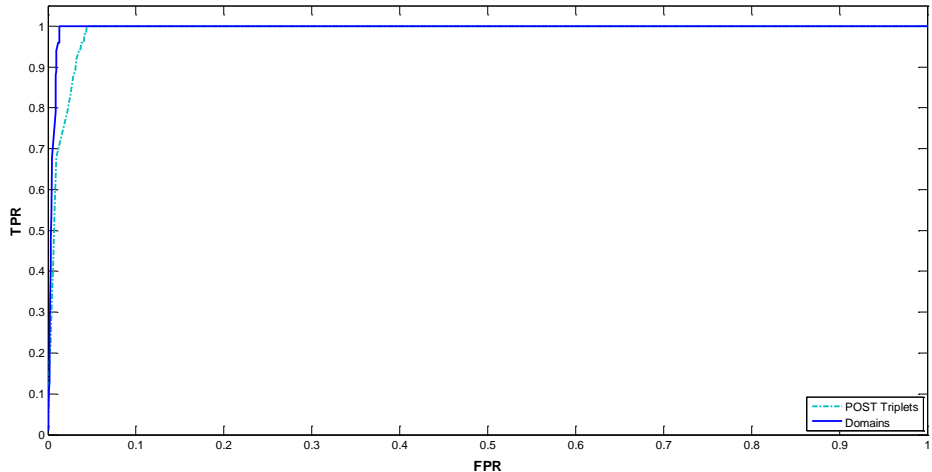

Fig. 13 Final ROC plots for CryptoWall 4.0 detection for merged Alexa and MACCDC datasets and after final fine-tuning of the detector.

**5. Conclusion**

In this paper we proposed a novel SDN-based ransomware detection system that is relying on the characteristics of malware communication. By performing network measurements for two very popular ransomware families i.e. CryptoWall and Locky we observed that a promising approach would be to detect malicious communication between an infected host and attacker C&C server using the HTTP traffic characteristics (HTTP messages sequences and their corresponding sizes). Therefore, we designed and developed a detection system that uses SDN solution to provide rapid reaction to the discovered threat. The experimental results obtained using real ransomware samples proved that even such a simple approach is feasible and offers good efficacy. We were able to achieve detection rate of 97-98% with 1-2% or 4-5% false positives when relaying on domains or POST triples respectively.

As to the countering ransomware, in general, we believe that it is vital to try to target and break the business model of the malware developers/exploiters by, for example, determining the most profitable malware families and disrupting their infrastructure. Complementary approach is to educate users to pay more attention on what type of files they open or what types of websites they visit. It is clear that less infections result in a lower profit and higher operation costs for the cybercriminals. Future work on ransomware detection involves taking into account also the sizes of the HTTP responses from the C&C server. Moreover, it is possible to combine the proposed approach with others e.g. those based on blacklisting of malicious IP addresses and domains. Furthermore, it is vital to monitor ransomware development trends in order to provide effective countermeasures as fast as possible to limit the number of infected users' machines.

**References**
[1] D. Kreutz, F. V. Ramos, P. Verissimo, C. Rothenberg, S. Azodolmolky, S. Uhlig. "Software-Defined Networking: A Comprehensive Survey", *Proc. of the IEEE*, 103(1):14-76, January 2015
[2] S. A. Mehdi, J. Khalid, S. A. Khayam, "Revisiting Traffic Anomaly Detection Using Software Defined Networking", *Proc. of the 14th International conference on Recent Advances in Intrusion Detection (RAID 2011)*, pp. 161-180, 2011
[3] A. Zaalouk, R. Khondoker, R. Marx, K. Bayarou, "OrchSec: An Orchestrator-Based Architecture For Enhancing Network-Security Using Network Monitoring and SDN Control Functions", *In Proc. of Network Operations and Management Symposium (NOMS),* pp. 1-9, 2014

[4] R. Jin, B. Wang, "Malware Detection for Mobile Devices Using Software-Defined Networking", *Proc. of GENI Research and Educational Experiment Workshop (GREE '13)*, pp. 81-88, 2013

[5] S. Shin, G. Gu, "CloudWatcher: Network security monitoring using OpenFlow in dynamic cloud networks (or: How to provide security monitoring as a service in clouds?)", *In Proc. of 20th IEEE International Conference on Network Protocols (ICNP)*, USA, 2012, pp. 1-6.

[6] Symantec, "Internet Security Threat Report", April 2015, URL: https://www4.symantec.com/mktginfo/whitepaper/ISTR/21347932_GA-internet-security-threat-report-volume-20-2015-social_v2.pdf

[7] Europol, "Internet Organised Crime Threat Assessment 2016 (iOCTA)", September 2016, URL: https://www.europol.europa.eu/content/internet-organised-crime-threat-assessment-iocta-2016

[8] McAfee Labs, "Meet 'Tox': Ransomware for the Rest of Us", May 2015, URL: https://blogs.mcafee.com/mcafee-labs/meet-tox-ransomware-for-the-rest-of-us/

[9] K. Savage, P. Coogan, H. Lau, "The evolution of ransomware, Symantec Security Response", August 2015, URL: http://www.symantec.com/content/en/us/enterprise/media/security_response/whitepapers/the-evolution-of-ransomware.pdf

[10] A. Kharraz, W. Robertson, D. Balzarotti, L. Bilge, E. Kirda, "Cutting the gordian knot: A look under the hood of ransomware attacks",*12th Conference on Detection of Intrusions and Malware & Vulnerability Assessment (DIMVA 2015)*, July 9-10, 2015, Milan, Italy

[11] K. Cabaj, P. Gawkowski, K. Grochowski, D. Osojca, "Network activity analysis of CryptoWall ransomware", *Przeglad Elektrotechniczny*, vol. 91, nr 11, 2015, ss. 201-204, URL: http://pe.org.pl/articles/2015/11/48.pdf

[12] Cuckoo Sandbox: Automated Malware Analysis, URL: https://cuckoosandbox.org

[13] Net Market Share, URL: https://www.netmarketshare.com/operating-system-market-share.aspx

[14] K. Cabaj, W. Mazurczyk, Using Software-Defined Networking for Ransomware Mitigation: the Case of CryptoWall, IEEE Network, November/December 2016, DOI: 10.1109/MNET.2016.1600110NM

[15] A. Tofilski, M. Couvillon, S. Evison, H. Helantera, E. Robinson, F. Ratnieks, Preemptive Defensive Self-Sacrifice by Ant Workers, In: American Naturalist, Vol. 172, No. 5, 11.2008, p. E239-E243.

[16] W. Mazurczyk, E. Rzeszutko, Security - a perpetual war: lessons from nature, IEEE IT Professional, vol. 17, no. 1, pp. 16-22, January/February 2015

[17] R. Perdisci, W. Lee, and N. Feamster, Behavioral Clustering of HTTP-Based Malware and Signature Generation Using Malicious Network Traces. In Proc. of the USENIX Symposium on Networked Systems Designs and Implementation (NSDI), April 2010.

[18] K. Rieck, G. Schwenk, T. Limmer, T. Holz, and P. Laskov, Botzilla: Detecting the Phoning Home of Malicious Software. In Proceedings of the 25th ACM Symposium on Applied Computing (SAC), March 2010.

[19] N. Idika, A. P. Mathur, A Survey of Malware Detection Techniques, Technical Report, Purdue University, 2007

[20] C. Rossow, C J. Dietrich, ProVeX: Detecting Botnets with Encrypted Command and Control Channels, In Proc. of 10th International Conference, DIMVA 2013, Berlin, Germany, July 18-19, 2013, pp. 21- 40

[21] Z. B. Celik, R. Walls, P. McDaniel, A. Swami, Malware traffic detection using tamper resistant features, Military Communications Conference (MILCOM), Tampa, FL, USA, October 2015

[22] R. Zegers, HTTP header analysis, M.Sc thesis, University of Amsterdam, 2015

[23] N. Kheir, Analyzing HTTP User Agent Anomalies for Malware Detection, In Proc. of 7th International Workshop, DPM 2012, and 5th International Workshop, SETOP 2012, Pisa, Italy, September 13-14, 2012, pp. 187-200

[24] M. Grill, M. Rehak, Malware Detection Using HTTP User-Agent Discrepancy Identification, IEEE International Workshop on Information Forensics and Security (WIFS), 2014, pp. 221-226

[25] J. M. Ceron, C. B. Margi, L. Z. Granville, MARS: An SDN-Based Malware Analysis Solution, IEEE Symposium on Computers and Communication (ISCC), Messina, 2016, pp. 525-530

[26] N. Andronio, Heldroid: Fast and Efficient Linguistic-Based Ransomware Detection, M.Sc. thesis, University of Illinois at Chicago, 2015

[27] D. Sgandurra, L. Muñoz-González, R. Mohsen, E. C. Lupu, Automated Dynamic Analysis of Ransomware: Benefits, Limitations and use for Detection, In: Computing Research Repository (CoRR), abs/ 1609.03020, arXiv.org E-print Archive, Cornell University, Ithaca, NY (USA), September 2016

[28] P. T. N, Scaife, H, Carter, K. R. Butler, Cryptolock (and drop it): Stopping ransomware attacks on user data. In 2016 IEEE 36th International Conference on Distributed Computing Systems, pp. 303-312, 2016.

[29] G. Gu, R. Perdisci, J. Zhang, W. Lee, Botminer: Clustering analysis of network traffic for protocol- and structure-independent botnet detection, In: Proceedings of the 17th USENIX Security Symposium, 2008

[30] P. Wurzinger, L. Bilge, T. Holz, J. Goebel, C. Kruegel, E. Kirda, Automatically Generating Models for Botnet Detection. In: Backes, M., Ning, P. (eds.) ESORICS 2009. LNCS, vol. 5789, pp. 232–249. Springer, Heidelberg (2009)

[31] M. Bailey, J. Oberheide, J. Andersen, Z. M. Mao, F. Jahanian, J. Nazario, Automated Classification and Analysis of Internet Malware. In: C. Kruegel, R. Lippmann, A. Clark (eds.) RAID 2007. LNCS, vol. 4637, pp. 178–197, 2007

[32] U. Bayer, P. M. Comparetti, C. Hlauschek, C. Kruegel, E. Kirda, Scalable, behavior-based malware clustering. In: Network and Distributed System System Security Symposium, 2009

[33] G. Jacob, R. Hund, C. Kruegel, T. Holz, Jackstraws: Picking command and control connections from bot traffic. In: 20th USENIX Security Symposium, 2011